\documentclass[12pt]{iopart}
\usepackage{graphicx} 
\usepackage{url}
\expandafter\let\csname equation*\endcsname\relax
\expandafter\let\csname endequation*\endcsname\relax
\usepackage{amsmath}
\usepackage{xcolor}
\usepackage{amssymb}
\usepackage{amsfonts}

\begin{document}

\title[Non-orbital particle trapping in binary black holes through dynamic stability]{Achieving non-orbital particle trapping in binary black holes through dynamic stability}

\author{A Kurmus$^{1,3}$, M Zaja\v{c}ek$^2$, G Kestin $^1$, L Deslauriers$^1$}
\address{$^1$ Department of Physics, Harvard University, Cambridge, MA 02138, USA}
\address{$^2$ Department of Theoretical Physics and Astrophysics, Faculty of Science, Masaryk University, Kotlářská 2, CZ-611 37 Brno, Czech Republic}
\address{$^3$ Department of Physics, University of Washington, Seattle, WA 98195-1560, USA}

\ead{louis@physics.harvard.edu}
\vspace{10pt}
\begin{indented}
\item[]November 2024
\end{indented}

\begin{abstract}
We present an interdisciplinary comparison between binary black hole systems and Radio Frequency (RF) Paul Traps, modeling the gravitational binary system as a rotating saddle near its center. This analogy connects these seemingly unrelated systems through the concept of dynamic stability. The rotating saddle potential is analytically tractable, allowing us to prove the existence of bounded charged particle trajectories under certain conditions. By focusing on stellar-mass black holes with a weak electric charge—a feature consistent with specific astrophysical conditions that leaves the spacetime metric largely unaffected but can influence nearby particle interactions—we can neglect complicating factors such as magnetic fields from large accretion disks of heavier black holes or stellar winds. Our simulation results demonstrate that charged particles can exhibit stable, non-orbital trajectories near the center of a binary system with charged stellar-mass black holes, providing unique three-dimensional trapping primarily through gravity. This system is distinctive in the literature for its non-orbital trapping mechanism. While theoretically intriguing, this trapping relies on specific conditions, including nearly identical black hole masses. These types of non-orbital trapping mechanisms could potentially allow for longer-lived plasma configurations, enhancing our ability to detect electromagnetic signatures from these systems. The significance of this work lies in the novel comparison between a laboratory-scale quantum system and a larger astrophysical one, opening new avenues for exploring parallels between microscopic and cosmic phenomena across fourteen orders of magnitude in distance.
\end{abstract}

%
%
\submitto{\CQG}
%
%
%

\section{Introduction} \label{sec:intro}

The development of advanced technologies for precise control and manipulation of atoms has had a profound impact on our understanding of quantum systems, leading to new applications. Among these technologies, Radio Frequency (RF) Paul Traps have emerged as crucial tools for manipulating atomic ions with exceptional precision. Devised by Wolfgang Paul, for which he received the Nobel Prize in Physics in 1989, RF Paul Traps use dynamic electric fields to confine charged particles in a region of space, enabling detailed studies on their properties and interactions \cite{Paul_1990}.

RF Paul Traps operate on the principle that it is impossible to confine particles in three dimensions using a static electric field due to Gauss's law. Instead, they achieve dynamic stability—a state where particles remain confined over time despite local instabilities—by producing a saddle potential and varying it over time. This approach allows for the confinement and manipulation of charged particles within specific spatial regions, facilitating numerous experimental endeavors such as high-precision measurements, quantum computing, and spectroscopy \cite{Deslauriers_2004,Leibfried_2003,Haffner_2008}.

This paper examines the application of dynamic stability principles to cosmic systems, specifically binary black holes. These systems, composed of two orbiting black holes, create a gravitational potential that changes over time, reminiscent of the time-varying potentials in RF Paul Traps. The gravitational potential of binary black holes can be understood as a rotating saddle potential, a concept long used as a mechanical analogy for RF Paul Traps. Despite the vast difference in scale over as much as fourteen orders of magnitude, the underlying physics of time-varying potentials in both systems suggests a compelling parallel worth investigating. This study aims to examine how the gravitational field of binary black hole systems could create regions of non-orbital trapping analogous to those found in Paul Traps. This novel trapping mechanism may provide new ways to think about matter accumulation in binary systems.

By focusing on stellar mass black holes, we simplify the comparison to a rotating saddle, avoiding complications from accretion disks, solar winds, and stellar magnetic fields that are more prominent in larger systems. This parallel between Paul Traps and rotating saddle potentials paves the way to explore gravitational analogs to Paul Traps in binary systems, particularly binary black hole systems that are made up of stellar mass black holes.

The stability of a particle's trajectory in a rotating saddle potential hinges on a delicate interplay between the potential's curvature and its rotation frequency. This dynamic stability can be understood as follows: at any given moment, the saddle potential creates a trapping force along one axis and an anti-trapping force along the perpendicular axis. As the potential rotates, these trapping and anti-trapping directions continuously shift. When the rotation frequency is appropriately tuned relative to the potential's steepness, a particle that begins to move away from the center along the momentary anti-trapping direction is swept into the trapping direction before it can escape. This constant redirection towards the center results in a bound orbit, effectively creating a dynamically stable trap.

In the context of binary black hole systems, however, achieving this stability through gravitational forces alone is impossible. This is due to Kepler's third law, which intrinsically links the curvature of the gravitational potential to the orbital frequency. As a result, the gravitational saddle potential is always too steep relative to its rotation frequency, placing it invariably in the unstable region. No matter how fast the system rotates, the curvature of the gravitational potential remains too sharp to allow for stable orbits.

This inherent instability in purely gravitational binary systems is what makes the recent findings about charged black holes noteworthy. Astrophysical research has revealed that black holes and their surrounding magnetospheres can possess a weak electric charge \cite{Neslusan_2001,2017FoPh...47..553E,Zajacek_2018,Zajacek_2019_b,Black_Hole_Pulsar, 2020ApJ...897...99T, Bozzola_PRL_2021,2024arXiv240917639N}, which, while small enough not to significantly affect the spacetime metric, can still influence nearby particle interactions, particularly the innermost stable circular orbit and overall stability \cite{Zajacek_2019_a,2022ApJ...937...50A}. This additional electromagnetic component reduces the net steepness of the potential relative to the rotation frequency, potentially allowing for stable trajectories that are otherwise impossible in purely gravitational systems. The mechanisms by which black holes acquire charge open new possibilities for gravitational-electromagnetic analogs to Paul Traps.

While the specific astrophysical relevance of these charging mechanisms remains uncertain, several viable processes have been proposed:
\begin{itemize}
    \item {Magnetically Induced Charging}:
    \begin{itemize}
        \item {Wald Mechanism}: A rotating (Kerr) black hole embedded in a large-scale poloidal magnetic field generates an electric field, leading to preferential charge accumulation \cite{Wald_1974,Zajacek_2018}.
        \item {Relative Motion}: A black hole moving through an ambient magnetic field, such as in a binary system, experiences Faraday induction, leading to charge separation and accumulation \cite{Adari_2023}.
    \end{itemize}    
    \item {Plasma and Accretion-Based Charging}:
    \begin{itemize}
        \item {Charge Imbalance in the Interstellar Medium}: Proton-electron separation in the gravitational field of a central mass \cite{Zajacek_2018,2024arXiv240917639N} or accretion of charged dust particles can result in a net charge \cite{2004NewAR..48..843E}.
        \item {Dark Matter Interactions}: Accretion of millicharged dark matter \cite{2018Natur.557..684M} or diffusion effects in virialized halos may allow black holes to acquire stable charge \cite{Araya_2023}.
    \end{itemize}    
    \item {Primordial and Evolutionary Charging}:
    \begin{itemize}
        \item {Primordial Black Holes}: Charge accumulation at formation due to Poisson fluctuations or high-energy particle interactions \cite{Araya_2023}.
        \item {Hawking Evaporation and Black hole Extremalization}: A weakly charged black hole may become extremal as it evaporates, with further evolution influenced by the weak gravity conjecture, potentially leading to charge redistribution between daughter black holes resulting from the split-up of the extremal charged black hole \cite{2007JHEP...06..060A,2007JHEP...12..068K,2018JHEP...10..004C}.
    \end{itemize}
\end{itemize}
These mechanisms involve charge with different degrees of stability and charging/discharging timescales, with charge accumulation competing against discharge processes. For this study, we assume that each black hole in the binary system holds a quasi-stable electric charge sufficient for particle trapping over at least several thousands of orbital periods. A more detailed discussion of these charging mechanisms and their astrophysical implications is provided in Appendix A.

The presence of charge in binary black hole systems creates conditions that may allow for particle confinement. The gravitational and electrostatic potential near the center of such a system approximates a rotating saddle. However, this approximation is not exact due to the nature of the potential created by two point sources. The actual potential deviates slightly from the ideal saddle form. Nevertheless, analyzing stability through the rotating saddle analogy provides a useful framework for understanding these cosmic trapping mechanisms. This approach allows us to apply established analytical techniques while accounting for the specific characteristics of binary black hole systems.

The comparison between charged binary black holes and Paul Traps, while unexpected, may lead to new insights into both trapping mechanisms and astrophysical phenomena. This interdisciplinary approach, bridging atomic physics and astrophysics, may offer unprecedented insights into the fundamental principles governing stability in vastly different scales of our universe.

In this paper, we systematically explore the possibility of gravitational-electromagnetic analogs to Paul Traps in binary black hole systems. Section \ref{sec:ideal_system} examines an idealized binary black hole setup that can be exactly represented as a rotating saddle, demonstrating the theoretical possibility of gravitational-electromagnetic trapping. Section \ref{sec:physical_system} extends this analysis to more realistic scenarios, showing that stable trajectories may still be possible when incorporating physical factors that cause deviations from the ideal rotating saddle potential. Section \ref{sec:plausibility} evaluates the astrophysical plausibility of conditions necessary for a binary black hole system to trap charged particles. Finally, Section \ref{sec:conclusions} summarizes our findings and discusses their implications for our understanding of both trapping mechanisms and cosmic phenomena.

\section{Modeling Binary Black Holes with a Rotating Saddle Potential}\label{sec:ideal_system}

This section analytically demonstrates the possibility of stable particle trajectories within an idealized binary black hole system. Our analysis relies on two key simplifications:

\begin{enumerate}
    \item We focus on the region near the center of the system, where the gravitational and electrostatic potential can be approximated as a rotating saddle. This approximation is crucial as it allows us to apply known analytical techniques for studying particle stability in saddle potentials.
    
    \item We simplify the system by assuming the black holes have constant electric charges and by ignoring any external magnetic field. The absence of a magnetic field is particularly important, as its presence would preclude modeling the overall potential as a saddle.
\end{enumerate}

These idealizations, while limiting physical plausibility, enable us to directly compare our system with rotating saddle potentials and provide a foundation for understanding more complex scenarios.

Our initial model consists of two identical black holes, each with mass $M$ and charge $Q$, orbiting each other, in a circular orbit with radius $b$ centered at origin $(0,0,0)$, in the $xy$ plane. The black holes are separated by a distance $2b$, with $\text{BH}_1$ initially located at $(b, 0, 0)$ and $\text{BH}_2$ at $(-b, 0, 0)$. The particle subjected to this potential has mass $m$ and charge $e$.

\subsection{Equations of Motion}\label{sec:eq_motions}

The analytical derivation of equations of motion for a particle within a charged binary black hole potential is challenging, particularly in the $x$ and $y$ directions. However, our analysis focuses on the region near the center of the system, where stable trajectories are theoretically possible. In this region, the potential approximates a perfect rotating saddle-like quadrupole potential in the $xy$ plane. While we begin with an idealized linearized potential for analytical tractability, we later verify in Section 3 that nonlinear effects remain small within the relevant parameter space, ensuring that our stability results remain robust.

This approximation allows us to conceptualize the system as a mechanism capable of achieving dynamic stability and trapping in the $xy$-plane, complemented by static trapping in the $z$ direction. By framing our setup as a rotating saddle, we can characterize it using the following potential:

\begin{equation}
\label{eq:saddle}
    V(x,y,z) \approx \frac{GMm-kQe}{b^3} \left(-\alpha x'^2 + y'^2 + z^2\right)
\end{equation}
where $G$ is the gravitational constant,  $k$ is the Coulomb constant, $Q$ is the black hole charge, $e$ is the particle charge, and $\alpha$ is the asymmetry parameter which, due to Gauss's law for gravitational and electric fields in three dimensions, will always be equal to $2$ in our setup. We retain $\alpha$ as a parameter to provide intuition for the stability parameters that we will introduce later and to make the comparison to a rotating saddle explicit. A detailed derivation of this rotating saddle potential from exact Newtonian forces can be found in \ref{ap_sec:potential_derivation}.

Additionally, the coordinates $x'$ and $y'$ are time-dependent $x$ and $y$ coordinates in the rotating frame of the saddle. We can express them in terms of $x$, $y$, and $t$ in the non-rotating frame as

\begin{equation}
\label{eq:primes}
    x' \rightarrow x\cos(\Omega t)+y\sin(\Omega t), ~\\
    y' \rightarrow y\cos(\Omega t)-x\sin(\Omega t)
\end{equation}
where $t$ represents time and $\Omega$ represents the frequency of the rotation of the saddle (which specifically, in our case, is the orbital frequency of the black holes). A key distinction between our modeled potential and a conventional rotating saddle lies in the fixed nature of the rotation frequency. While $\Omega$ can take on any value in a rotating saddle, it is determined by Kepler's third law in a binary black hole system. Applying Kepler's Law we find that
\begin{equation}\label{eq:omega}
    \Omega_{\text{grav}} = \sqrt{\frac{GM}{4b^3}}.
\end{equation}

For dynamic stability, the rotation frequency of the saddle $\Omega$ must be significantly larger than the characteristic frequency of particle motion in the trap. Specifically, stability requires that the orbital frequency of the black holes $\Omega_{grav}$ be at least twice as large as the characteristic oscillation frequency of the trapped particle. This ensures that the particle does not escape before the effective potential can restore it, similar to how stability is maintained in Paul traps. In the following sections, we introduce the stability parameters $a$ and $q$, which formally encapsulate this condition and govern the particle’s motion in the rotating saddle potential.

With all variables now defined, we can simplify our analysis by decomposing the three-dimensional problem into two more manageable components: 
1) solving for the static potential in the $z$ direction, and 
2) addressing the rotating saddle potential in the $xy$ plane. 
To make notation easier, we define the variable $K$ as
\begin{equation}
    K  = \frac{GMm-kQe}{b^3}
\end{equation}
Then, inserting the expressions in Eq.~\ref{eq:primes} into Eq.\ref{eq:saddle} and taking the negative gradient of the potential we can get differential equations of motions for three dimensions as:
\begin{itemize}
    \item for $x$ direction
    \begin{equation}
    m \frac{{d^2 x}}{{d t^2}} + xK(\alpha - 1) + 2K\left(\frac{{\alpha + 1}}{2}\right)[x\cos(2\Omega t) + y\sin(2\Omega t)] = 0
\end{equation}
    \item for $y$ direction
    \begin{equation}
      m \frac{{d^2 y}}{{d t^2}} + yK(\alpha - 1) - 2K\left(\frac{{\alpha + 1}}{2}\right)[y\cos(2\Omega t) - x\sin(2\Omega t)] = 0
\end{equation}
    \item for $z$ direction
    \begin{equation}
         m \frac{{d^2 z}}{{d t^2}} + 2Kz = 0
    \end{equation}
\end{itemize}

From these equations, we can readily see that stable trapping occurs in the $z$ direction. However, determining stability in the $xy$ plane is less straightforward due to the complex form of the equations. To facilitate our analysis of motion in the $xy$ plane, we introduce the following change of variables:

\noindent Letting $\tau = \Omega t$, $a = -\frac{K(\alpha - 1)}{m\Omega^2}$, and $q = \frac{K(\alpha + 1)}{2m\Omega^2}$ yields
\begin{equation}\label{eq:mathieu_x}  
    \frac{d^2 x}{d \tau^2} - [a + 2q\cos(2\tau)]x - 2q\sin(2\tau)y = 0 
\end{equation}
and
\begin{equation}\label{eq:mathieu_y}
    \frac{d^2 y}{d \tau^2} - [a - 2q\cos(2\tau)]y - 2q\sin(2\tau)x = 0
\end{equation}

The stability of trajectories described by the modified Mathieu equations above can be determined through the dimensionless parameters $a$ and $q$. While the formal Mathieu equations describing Radio Frequency (RF) Paul Traps do not possess analytical solutions, the cross terms in our equations make an analytical solution possible. We can solve these equations by promoting them to the complex plane (defining $z = x + iy$ and using a rotating wave approximation \cite{Thompson_CJP_2002}). This approach yields an analytical solution, revealing that bounded solutions are only possible if $a \geq 2q + 1$, or when $a$ lies in the interval $[-q^2, -2q + 1]$. Explicit derivation of the stability regions are available in \ref{ap_sec:stab_eq_derivation}. The stability parameters $a$ and $q$ offer insights into the forces at play in our system. The parameter $a$, related to the asymmetry in the saddle (when $\alpha \neq 1$), can be interpreted as representing a static force. A perfectly symmetric saddle corresponds to $a = 0$. The parameter $q$, which relates to the curvature of the saddle potential, can be understood as representing a ponderomotive-like restoring force. We stress that while $a$ is influenced by $\alpha$, it also incorporates the effect of rotation frequency, thus representing the overall asymmetry in the rotating binary black hole system.

\subsection{Analytical Stability Parameters of the Rotating Saddle Model}\label{sec:stability_param}

\begin{figure*}
    \centering
    \includegraphics[width=\textwidth]{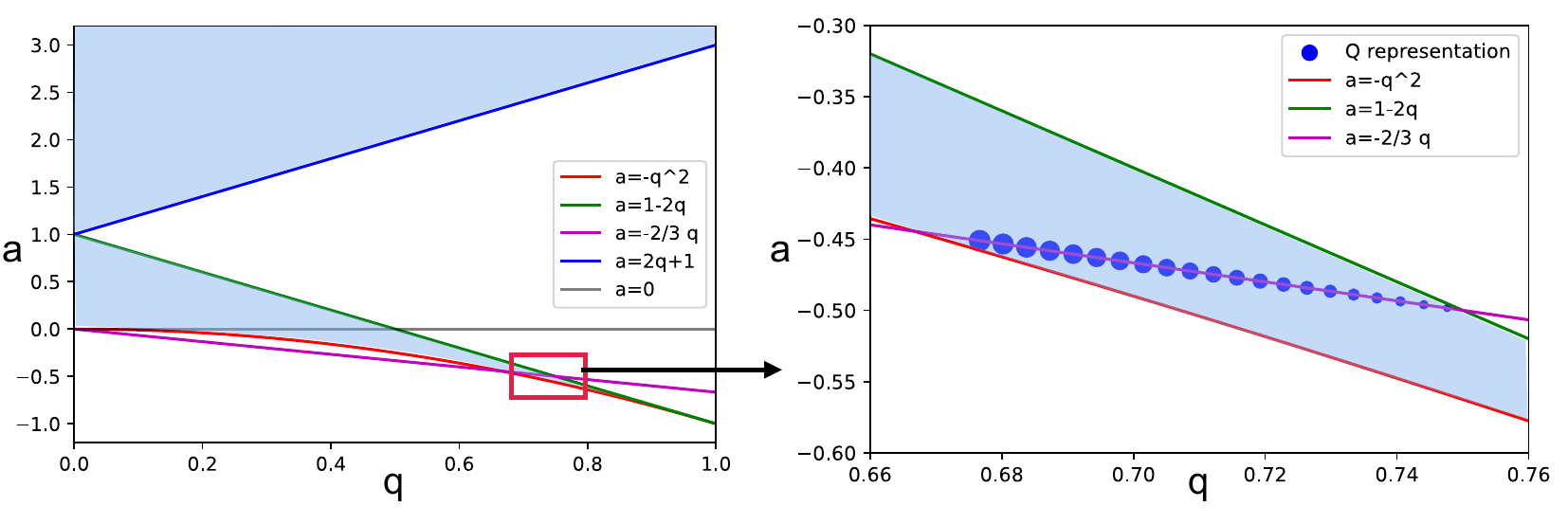}
    \caption{Regions of stability in the $(q, a)$ plane for a rotating saddle and the physically realizable region in a binary black hole potential. The horizontal and vertical axes in both panels represent the stability parameters $q$ and $a$, respectively. The left panel shows the entire stability region of particles whose trajectory is governed by the modified Mathieu equations in Eq.~\ref{eq:mathieu_x} and Eq.~\ref{eq:mathieu_y}. All regions shaded in light blue are regions of stability. The region identified with a red square shows where a stable trajectory in a binary black hole potential will be located within the entire range of stability. The right panel shows the zoomed-in version of the red square. The red curve and green line indicate the limits of general stability in the zoomed region. The magenta line represents the constraint due to the fixed asymmetry parameter $\alpha = 2$ enforcing the relation $a = -2/3q$. The blue points on the magenta line illustrate how varying the black hole charge $Q$ between 132.2 and 134.2 Coulombs affects the location of the stability line, corresponding to black holes of 1 solar mass. A larger data point represents a larger $Q$.}
    \label{fig:stability_diagram}
\end{figure*}

The region of stability in the $(q,a)$ plane is defined by $a \geq 2q + 1$, or when $a$ lies in the interval $[-q^2, -2q + 1]$, as illustrated by the blue-shaded area in Fig.~\ref{fig:stability_diagram}. However, the physical constraints of our binary black hole system limit the available stability region. Specifically, trapping in the $z$ direction requires that the gravitational force exceed the electrostatic force, which eliminates the possibility of $a > 0$ in our system. This is because $a > 0$ would imply a net repulsive force, making trapping in all three dimensions impossible. Furthermore, the quadrupole potential in our system dictates that the asymmetry parameter $\alpha$ must equal 2. This constraint, combined with the requirement for $z$-direction trapping, enforces a strict relationship between $a$ and $q$:

\begin{equation}
a = -\frac{2}{3}q.
\end{equation}

\noindent This relationship further narrows the region of stability available to our system, as shown by the magenta line in Fig.~\ref{fig:stability_diagram}.

Despite these constraints, there exist physically realizable combinations of black hole and particle properties that satisfy the stability conditions. Fig.~\ref{fig:stability_diagram} depicts the compact parameter space of stability for a system with positively charged black holes, which are motivated by the analysis of the charge of astrophysical black holes due to symmetries  \cite{Zajacek_2018,2020ApJ...897...99T}. To illustrate, consider a proton in the potential of two solar-mass black holes ($M=M_{\odot}$) separated by 2 AU, each with a charge of $Q = 133.2$ Coulombs. This configuration yields stability parameters of $a=-0.47$ and $q=0.705$, which fall within the permissible region. Fig.~\ref{fig:stability_diagram} also demonstrates how varying the black hole charge $Q$ affects the stability parameters. As $Q$ increases, $q$ decreases while $a$ increases. For this particular configuration, stability persists for Q values ranging from 132.2 to 134.2 Coulombs, highlighting a narrow but existent window of stability for 1 solar mass black holes. This range scales proportionally with black hole mass.

In conclusion, our analytical calculations demonstrate the theoretical possibility of particle trapping within this simplified model of a charged binary black hole system. This novel result opens up avenues for exploration, particularly in the context of the early Universe. During cosmic epochs when the ambient cosmological magnetic field was weaker (below $10^{-12}$ Gauss), our model's neglect of magnetic fields becomes a reasonable approximation. Moreover, various mechanisms for black hole charging exist beyond the Wald mechanism, which requires a magnetic field. These include Poisson fluctuations at the horizon (\cite{Araya_2023}; see also Sec.~\ref{sec:intro}), further broadening the applicability of our model. Consequently, our findings have particular relevance in the study of charged primordial black holes in the early Universe, where conditions may have been more conducive to the scenarios explored here. In case dark matter would at least partially be contributed by primordial black holes, it is plausible that these black holes formed binary systems with a certain fraction of nearly equal-mass binaries depending on the primordial black-hole mass distribution.

We note that in addition to neglecting the background magnetic field, we also neglect the presence of nonlinearities when we argue that the binary system reduces to a rotating saddle near the center. While these simplifications are necessary for our analytical approach, they raise questions about the model's applicability to more realistic scenarios. To address these concerns and validate our findings, we turn to numerical simulations in the following section. As we will show in Sec.~\ref{sec:physical_system}, these simulations incorporate both the background magnetic field and nonlinear effects, demonstrating that these perturbations do not prevent the possibility of trapping under certain conditions.

\section{Stable Trajectories In Physical Binary Black Hole Systems}\label{sec:physical_system}

In this subsection, we take into consideration the effects neglected in the theoretical analytical treatment, namely background magnetic field and nonlinear effects.

The introduction of a magnetic field adds a Lorentz force to the particle dynamics, alongside the existing gravitational and electrostatic forces. This addition results in a velocity-dependent term in the equations of motion. Specifically, the cross product of velocity and the magnetic field in the Lorentz force makes the $y$ velocity of the particle crossed with magnetic field in $z$ direction affect the equations of motion in the $x$ direction. Incorporating this term and transitioning from $t$ to $\tau$, we can modify Eq.\ref{eq:mathieu_x} and Eq.\ref{eq:mathieu_y} as follows:

\begin{equation}\label{eq:mag_x}
    \frac{d^2 x}{d \tau^2} - \left[a + 2q \cos(2\tau)\right] x - 2q \sin(2\tau) y + \frac{\omega_o}{\Omega} \frac{dy}{d\tau} = 0
\end{equation}
and
\begin{equation}\label{eq:mag_y}
    \frac{d^2 y}{d \tau^2} - \left[a - 2q \cos(2\tau)\right] y - 2q \sin(2\tau) x - \frac{\omega_o}{\Omega} \frac{dx}{d\tau} = 0
\end{equation}. 

Here, $\omega_o = eB/m$ represents the cyclotron frequency, which characterizes the circular motion of a charged particle in a uniform magnetic field. It is expressed in terms of the particle's charge $e$, mass $m$, and the magnitude of the magnetic field $B$. For our analysis, we assume that $B$ is perpendicular to the rotation plane (i.e., aligned with the $z$-axis). This orientation maximizes the Lorentz force's effect on the particle's motion in the $xy$ plane, allowing us to examine the most significant potential disruption to the trapping mechanism.

The addition of the magnetic field complicates our analysis in two main ways. First, it introduces new dynamics that make direct comparison with a rotating saddle potential less straightforward. Second, the velocity-dependent term in the equations of motion renders them analytically unsolvable, necessitating the use of numerical methods. Despite these complications, the stability parameters described in Sec~\ref{sec:stability_param} still provide valuable intuition about the system's behavior.

To address these challenges and gain a more comprehensive understanding of the system, we turn to numerical simulations. These methods not only allow us to solve the modified equations of motion but also enable us to model particle trajectories under more realistic conditions, including nonlinear effects. Our approach involves two key steps:

\begin{enumerate}
    \item We calculate the exact force on a particle due to both the black holes and the magnetic field, accounting for all relevant interactions.
    
    \item Using these force calculations, we employ a fourth-order Runge-Kutta method (RK4) to numerically integrate the equations of motion and determine the particle's trajectory over time.
\end{enumerate}

This numerical approach allows us to explore the system's behavior beyond the limitations of our analytical model, providing insights into the stability and dynamics of charged particles in this complex gravitational and electromagnetic environment.

We calculate the net force on the particle as a vector:
\begin{equation}
\mathbf{F}_{\text{net}} = (F_{1px} + F_{2px} + v_yB, F_{1py} + F_{2py} - v_xB, F_{1pz} + F_{2pz})
\end{equation}
where $F_{[1,2]p[x,y,z]}$ denotes the force on the particle due to the black holes, $v_x$ and $v_y$ are the velocity components of the particle in the $x$ and $y$ directions, and $B$ is the magnetic field strength in the $z$ direction.
The gravitational and electrostatic force from each black hole takes the form:
\begin{equation}
F_{1px} = \frac{|GMm-kQe| \cdot (x_{b1} - x_p)}{((x_{b1} - x_p)^2 + (y_{b1} - y_p)^2 + (z_p)^2)^{\frac{3}{2}}}
\end{equation}
where $x_p$, $y_p$, $z_p$ are the particle's coordinates; $x_{b1}$, $y_{b1}$, $x_{b2}$, $y_{b2}$ are the $x$ and $y$ positions of the first and second black hole; $M_1 = M_2 = M$ and $Q_1 = Q_2 = Q$ are the masses and charges of the black holes.
We relate the black hole positions to the time evolution parameter $\tau$ as:
\begin{equation}
(x_{b1}, y_{b1}) = (b \cos(\tau), b \sin(\tau))
\end{equation}
\begin{equation}
(x_{b2}, y_{b2}) = (b \cos(\tau + \pi), b \sin(\tau + \pi))
\end{equation}
where $b$ is the radius of the black holes' orbit.
These calculations allow us to precisely model the particle's trajectory in our simulations, accounting for both gravitational and electromagnetic effects in the binary black hole system.

\begin{figure}
    \centering
    \includegraphics[width=0.6\columnwidth]{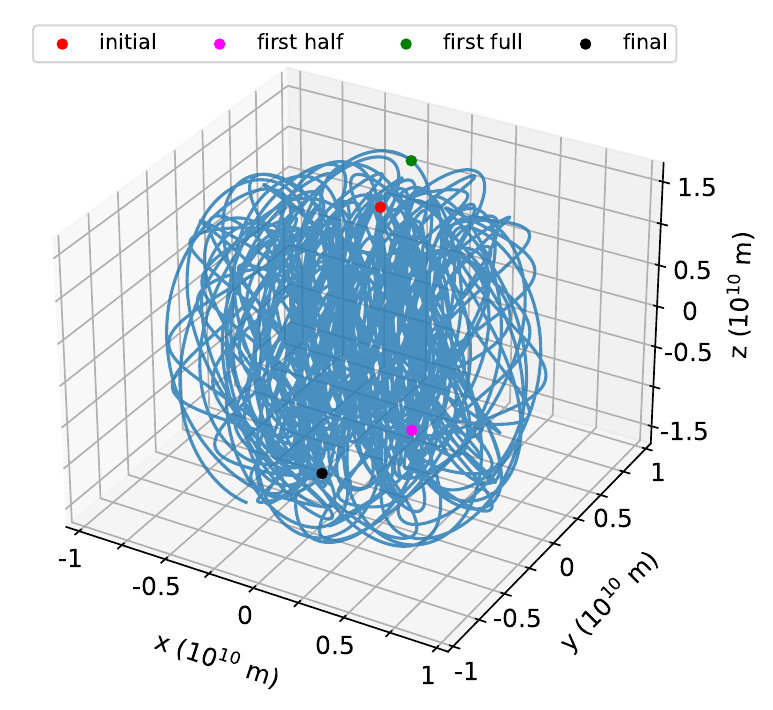}
    \caption{Trajectory of a proton under a charged binary black hole system potential with a background magnetic field of $10^{-13}$ Gauss determined by simulation. The axes labeled as $x$, $y$, and $z$ show the particle's position in units of $10^{10}$ m (note: 1 AU $= 1.496 \times 10^{11}$ m). The blue solid line represents the particle's trajectory for 100 full rotations of the binary system, equivalent to approximately 200 years for two solar-mass black holes ($M = \rm{M_{\odot}}$) separated by 2 AU. Single points indicate the particle's position at the start, after half a rotation, after a full rotation, and at the end of the plotted trajectory. While we evolved the simulation for $10^5$ rotations ($2 \times 10^5$ years), we display only the first 100 rotations for visual clarity. This figure demonstrates that a proton in a charged binary black hole potential can maintain a stable trajectory even with a background magnetic field, given appropriate combinations of black hole mass, charge, and magnetic field strength.}
    \label{fig:trajectory_clean}
\end{figure}

While simulations allow us to explore trajectories for various combinations of particle and black hole properties, specific parameter choices are crucial for this study. As discussed in Sec.~\ref{sec:stability_param}, only certain parameter combinations yield analytically stable trajectories. Moreover, to maintain the analogy with a Paul Trap and a rotating saddle, the cyclotron frequency induced by the magnetic field must be significantly lower than the black holes' orbital frequency. This ensures that the velocity-dependent terms in Eq.~\eqref{eq:mag_x} and Eq.~\eqref{eq:mag_y} remain negligible compared to other terms. To treat the magnetic field effect as a perturbation rather than a dominant force, we limited our analysis to fields up to $10^{-12.5}$ Gauss. A proton's trajectory in the presence of two solar-mass black holes ($M = \rm{M_{\odot}}$), each with a charge of 133.4 Coulombs ($Q = 133.4$ C), and a background magnetic field of $10^{-13}$ Gauss, satisfies these conditions. Figure~\ref{fig:trajectory_clean} illustrates the simulated trajectory of a proton under these conditions, suggesting the existence of a stable orbit. This is consistent with our analytical argument for stability in Sec.\ref{sec:ideal_system}. Using the same physical setup that yields the stability parameters ``$a$'' and ``$q$'', we find that the velocity-dependent term introduced by the magnetic field has a prefactor of $\omega_o / \Omega_{grav}$, making it at least two orders of magnitude smaller than the dominant terms in the ideal setup. Additionally, we show that the effect of nonlinearities remains perturbative within the simulated range (i.e., for particles within a radius of approximately $0.1b$ from the center), as detailed in \ref{ap_sec:potential_derivation}. In case the orbital frequency is larger than studied here, e.g. when the components are closer or they are more massive for a given distance, the magnetic field could be proportionally higher.

In our simulations, we use the number of black-hole binary rotations as a measure of time and the fraction of orbit radius, $b$, as a measure of size for stability. For two solar-mass black holes separated by 2 AU, a full rotation takes approximately $P_{\rm orb}\simeq 2(b/{\rm 1\,AU})^{3/2}(M/1\,M_{\odot})^{-1/2}$ years. We simulate the trajectory of a proton with an initial position of $(0.01b, 0, 0.1b)$ and a zero initial velocity. We evolve the simulation for $10^5$ black-hole binary rotations, corresponding to $2 \times 10^5$ years. The simulation results suggest that the particle is confined within a radius of $0.067b$ to $0.1b$ in the $xy$ plane and within $0.1b$ in the $z$ direction. Based on additional simulations, we qualitatively observe that the region of stability in the $xy$ plane is approximately circular, with a radius of about $0.1b$ (10\% of the black hole orbit radius).

Figure~\ref{fig:trajectory_clean} shows the trajectory of the particle up to 100 rotations and denotes the final location of the particle after $10^5$ rotations. We observe that even after $10^5$ rotations, the particle remains within the region it occupied during the first 100 rotations. The particle does not leave the compact region with a radius of approximately $0.1b$ (equivalent to 0.1 AU in our example) during multiple black-hole binary orbits, indicating that it is indeed physically confined with a stable trajectory. This demonstrates that the charged binary black-hole system is capable of trapping charged particles in a manner phenomenologically analogous to a Paul Trap, with simulations confirming that both nonlinearities in the gravitational and electric fields, as well as the weak ambient magnetic field, have a negligible impact on stability within this trapping volume.

Importantly, we obtain identical simulation results when using other combinations of black hole and magnetic field properties that yield the same stability parameter values ($a=-0.47$ and $q=0.705$) and maintain the same ratio between cyclotron motion and binary black-hole orbital motion frequencies. For instance, the black hole orbital frequency is proportional to black hole mass, while the cyclotron frequency is proportional to magnetic field strength. If we increase the black hole mass and charge by a factor of 10 and the magnetic field strength by a factor of $\sqrt{10}$, we obtain an identical trajectory. This theoretical consistency validates the output of our simulations (see Appendix D for a detailed discussion of these scaling relationships).

While further exploration could deepen our quantitative understanding of trapping regions and other dynamics, such inquiries hinge on the plausibility of the physical parameter combinations. We address the likelihood of these conditions in Section~\ref{sec:plausibility}.

\section{Astrophysical Constraints on Stable System Parameters}\label{sec:plausibility}

Our analysis demonstrates that within a charged binary black hole system's potential, even in the presence of a background magnetic field, a charged particle (such as a proton) can maintain a stable trajectory comparable to one exposed to a rotating saddle potential, and consequently, a Paul Trap. The black holes in our model are of solar mass order, a plausible range for both stellar and primordial black holes \cite{Domingo_2021}. For such black holes, accretion disks are sufficiently small to be negligible in our model. This assumption is supported by the non-detection of electromagnetic counterparts for LIGO/VIRGO merging binaries. Furthermore, the concept of primordial black holes contributing to dark matter implies negligible electromagnetic emission from accretion. The plausibility of a charge of 133.4 C/$M_{\odot}$ is supported by the range of charges for primordial black holes, between $0.1$ C/$M_{\odot}$ and $10^{15}$ C/$M_{\odot}$ \cite{Araya_2023}. Although the value of $Q\simeq 133$ C may seem to be contrived and fine-tuned, it is of the same order of magnitude as the value expected for any macroscopic object embedded in fully ionized plasma \cite{Neslusan_2001,1978ApJ...220..743B} to offset the separation of lighter electrons from heavier protons. The Wald mechanism \cite{Wald_1974} can lead to similar or even larger values of charge, depending on the spin and the strength of the external magnetic field. Other mechanisms relevant for primordial black holes such as Poisson fluctuations in the number density of particles during the horizon crossing in combination with the shielding from discharge provided by the magnetic field of maximally rotating primordial black holes can lead to a large range of charge-to-mass ratios specified earlier that can be stable over cosmological periods of time \cite{Araya_2023}.
Additionally, our model's background magnetic field strength, ranging from $10^{-12}$ to $10^{-13}$ Gauss, aligns with existing literature on magnetic field strength in the early Universe \cite{Ratra_1992,2004PhRvD..70l3003B}.

The theoretical and simulation analyses presented above explored scenarios where both black holes had identical mass and charge. However, in a cosmic context, finding such a perfectly matched binary system is rare, though not impossible given the vast number of binary black hole systems in the universe. Introducing a mass difference between the black holes causes the stability region to rotate with the black holes, rather than remaining stationary in space. This rotation introduces phenomena like parametric resonance, which can destabilize particle trajectories.

To quantitatively assess the effects of mass differences, we conducted additional simulations with varying black hole masses. We characterized the mass difference using the parameter $\mu$, defined as the ratio of the more massive black hole to the total system mass: $\mu = \frac{M_1}{M_1 + M_2}$. We assumed that black hole charges scaled proportionally with mass, maintaining the condition $\frac{M_1}{Q_1} = \frac{M_2}{Q_2}$. Our simulations revealed that stable trajectories become unfeasible when the mass ratio $\mu$ exceeds 0.505, corresponding to a maximum mass difference of approximately 1\% between the two black holes.

While this constraint on mass differences narrows the range of systems capable of exhibiting this trapping behavior, it does not preclude their existence entirely. Interestingly, observations of binary stellar-mass black hole mergers reveal a mass ratio distribution that follows a decaying power law, with a peak near equal masses \cite{Tiwari_2021}. In addition, in case primordial charged black holes were formed in the early Universe, they likely formed binary systems with a certain fraction of nearly equal-mass binaries that increases for approximately flat mass distributions or distributions with low and high mass cut-offs. This observed tendency towards similar-mass pairings in binary black hole systems aligns with the conditions required for our proposed trapping mechanism. Although such precisely matched systems may be uncommon, the vast number of binary black holes in the universe suggests that these gravitational analogs to Paul Traps could exist in nature, offering possibilities for future research and observation.

For specific masses of nearly equal binary black hole components, there is a narrow range of charge that would enable stable particle trapping. On one hand, this limits the broader realization in nature. On the other hand, the mechanism is well defined and the stability of particle trapping is set by specific conditions. Hence, this implies that the mechanism can be realized in nature when the right conditions are met. Although the narrow conditions limit the relevance from a quantitative (statistical) point of view, the plausible set-up can be studied in detail in the laboratory using the RF Paul trap analog. This is not the case for many other astrophysical phenomena that are expected to be generally applicable (thermonuclear reactions in stars). Therefore, the mechanism provides a unique opportunity to study the non-orbital particle trapping in binary black hole systems in a cross-disciplinary way. Its relevance for e.g. the evolution of primordial black holes and their interaction with the surrounding medium can thus further be assessed. The stable particle trapping by a near-equal mass black hole binary requires sufficiently stable electric charge, e.g. at least during several orbits, and the magnetic field that is weak enough so that the cyclotron frequency is lower than the orbital frequency.

The primordial black hole binary setup studied in the paper and its capability of stable trapping of particles may be of relevance for the studies of their formation and evolution, which is relevant for the structure formation in case they would contribute partially to the dark sector of the universe \cite{2020PhRvD.101d4022E,2021MNRAS.507.4804S}. In addition, the particle accumulation near their center of mass could contribute to matter-energy inhomogeneities and their understanding in the early universe when primordial black holes were forming. Furthermore, the trapping of particles could be relevant to the ``weak gravity conjecture’’ models \cite{2007JHEP...06..060A,2007JHEP...12..068K} that predict the splitting of extremal black holes with $Q=M$ into the pair of two ``daughter” black holes of potentially nearly equal mass and charge. The model could be extended to current black hole pairs in low-density environments with a weak magnetic field. For instance, in the intergalactic space, the particle densities can reach $10^{-5}\,{\rm cm^{-3}}$ and magnetic field is of the order of $10^{-9}$ G \cite{2001PhR...348..163G,2020rfma.book..279K}. In addition, over the Hubble time, black holes within dark-matter halos could charge via diffusion \cite{Araya_2023}. For such conditions, a pair of black holes should be separated by $\lesssim 0.004$ AU so that the particle trapping is stable. In general, for most charging mechanisms to work and to trap other free charged particles, presence of freely moving charge particles is necessary, which limits the applicability of the model to the early times before the proton-electron recombination, $z>1100$, and subsequently to the times at and after the hydrogen reionization, $z\lesssim 30$.

\section{Conclusion}\label{sec:conclusions}

This paper explored the concept of a gravitational analog to a Paul Trap by analyzing charged binary black hole systems. We established this parallel by connecting Paul Traps to rotating saddle-like potentials, both of which provide dynamic stability. We further demonstrated that a binary black hole potential reduces to a rotating saddle potential in regions near the system's center of mass. Building on this connection, we theoretically demonstrated the potential for stability within a rotating saddle potential, which approximates the behavior of a charged binary black hole system near its center. We verified this stability hypothesis through simulations that incorporated nonlinearities and background magnetic fields, elements not considered in the theoretical model.

Our initial theoretical predictions and simulations explored stability in an idealized setup with equal masses and charges for both black holes. To assess the physical feasibility of gravitational trapping, we introduced mass differences between black holes in our simulations. These revealed that even a 1\% variance can lead to instability, suggesting that the conditions for such trapping are highly specific. Despite the narrow parameter range for stability, the concept remains astrophysically relevant. Observed merging black-hole pairs tend to have mass ratios near unity, aligning with our stability requirements. Moreover, several standard and non-standard astrophysical and cosmological models consider black holes (both stellar and primordial) acquiring stable electric charges.

The most promising analogs to Paul Traps in our universe could be pairs of nearly equal-mass primordial, charged black holes in the early Universe, when the background magnetic field was sufficiently small. This scenario offers an intersection of quantum mechanics, general relativity, and astrophysics. In light of these findings, our work demonstrates the phenomenological possibility of gravitational, non-orbital trapping of particles within binary black hole systems. Recent analyses supporting the possibility of black holes possessing stable electric charges further strengthen this analogy to artificial Paul Traps.

Specific conditions for trapping limit the broad astrophysical applicability of the phenomenon. In addition, even if such an equal-mass, charged pair has existed, the stability of particle trapping is limited by discharge mechanisms. However, several mechanisms allow for stable charge, which, on the other hand, would significantly enhance the relevance of this process, especially taking into account a potentially large number of primordial black holes. Given these uncertainties, it is challenging to assess the full astrophysical as well as cosmological relevance of the process at the moment, but it certainly deserves attention due to the direct connection with controlled laboratory experiments with RF Paul traps.

While the precise conditions required for such trapping may be uncommon in the current universe, the vast number of black hole systems suggests that these phenomena could exist considering the whole cosmic history. More importantly, this research opens up possibilities for further exploration. It invites us to consider what other analogs of artificial systems might exist within the cosmos, potentially bridging the gap between laboratory physics and astrophysical phenomena. This work not only contributes to our understanding of charged black hole systems but also exemplifies how concepts from different areas of physics can yield unexpected insights when combined. As we continue to explore the universe, such interdisciplinary approaches may reveal more surprising connections between the physics of the very small and the very large.

\ack
 We acknowledge significant contributions from Aidan Carey and Laurel Barnett; along with valuable discussions with Daniel Davis, Anna Klales and Logan McCarty. We gratefully acknowledge support from the Faculty of Arts and Sciences at Harvard University. GPT-4 \cite{OpenAI2023ChatGPT4} was used to refine this manuscript. MZ acknowledges the GA\v{C}R Junior Star grant no. GM24-10599M for support. We thank the referees for making a number of helpful suggestions that have improved the clarity of our work.

\section*{Data Availability}

The data presented in the figures are available upon request from the corresponding author.

\section*{References}


\bibliography{references}  


\appendix
\section{Astrophysical Mechanisms for Black Hole Charging}

In this appendix, we discuss the primary astrophysical mechanisms through which black holes can acquire and maintain a weak but stable electric charge. These mechanisms, while differing in their physical origins, all contribute to the possibility of charged binary black holes playing a role in gravitational-electromagnetic trapping.

\subsection{Magnetically Induced Charging}

One of the most well-understood mechanisms for black hole charging involves interactions with external magnetic fields:

\begin{itemize}
    \item \textbf{Wald Mechanism}: A rotating (Kerr) black hole immersed in an external poloidal magnetic field develops an electrostatic potential difference between the event horizon and infinity. This potential difference results in the accumulation of a net charge \cite{Wald_1974,Zajacek_2018} that can be both positive or negative. The maximum charge attained is given by the Wald charge:
    \begin{equation}
        Q_{\rm W} = 2 a M B_{\rm ext}\lesssim 145 \left(\frac{M}{1\,M_{\odot}} \right)^2 \left(\frac{B_{\rm ext}}{10\,{\rm G}} \right)\,{\rm C},
    \end{equation}
    where $a$ is the dimensionless spin parameter, $M$ is the black hole mass, and $B_{\rm ext}$ is the strength of the external magnetic field. The limit on the right corresponds to the maximum rotation parameter $a\lesssim M$ corresponding to an extremal black hole.
    \item \textbf{Motion-Induced Charging}: Black holes moving relative to an external magnetic field (such as in a binary system) experience Faraday induction, leading to charge separation. This process occurs due to the generation of an electric potential between different regions of the black hole, analogous to a conducting sphere moving in a magnetic field \cite{Adari_2023}.
\end{itemize}

\subsection{Plasma and Accretion-Based Charging}

Beyond interactions with the surrounding magnetic field, black holes may also acquire charge from their surrounding environment, including interstellar plasma and accretion processes:

\begin{itemize}
    \item \textbf{Charge Imbalance in the Interstellar Medium}: In astrophysical environments, gravitational potential wells of black holes can induce proton-electron separation due to the mass difference. This separation is halted by the induced charge on the order of \cite{Neslusan_2001,1978ApJ...220..743B,Zajacek_2018,2024arXiv240917639N}
    \begin{equation}
        Q_{\rm eq} = \frac{2 \pi \epsilon_0 G (m_{\rm p}-m_{\rm e})}{e} M   \sim 77\,\left(\frac{M}{1\,M_{\odot}} \right){\rm C} 
    \end{equation}
     
    where $\epsilon_0$ is the vacuum permittivity, $m_{\rm p}$ and $m_{\rm e}$ are the proton and the electron masses, respectively, and $e$ is the electron charge. Alternatively, if a black hole moves through an irradiated dust cloud, positively charged dust grains may preferentially accrete onto it due to the Bondi-Hoyle-Lyttleton accretion \cite{2004NewAR..48..843E}, leading to a net positive charge . Conversely, within virialized dark-matter halos, black holes can acquire negative charge via diffusion due to the higher thermal speeds of electrons compared to ions \cite{Araya_2023}.

    \item \textbf{Interactions with Dark Matter}: In some models of beyond-standard-model physics, dark matter may carry small electric charges (``millicharged" dark matter \cite{2018Natur.557..684M}). If such particles interact with black holes, they could contribute to long-term charge accumulation \cite{Cardoso_2016}.
\end{itemize}

\subsection{Primordial and Evolutionary Charging Mechanisms}

Additional charge acquisition mechanisms are relevant for primordial black holes and those undergoing long-term evolution:

\begin{itemize}
    \item \textbf{Primordial Black Hole Formation}: Charge may be imprinted on black holes at formation due to statistical Poisson fluctuations in the number density of particles at horizon crossing. This is based on the purely statistical argument that there were patches of space with a small charge overabundance in the early universe that were not in causal contact. This mechanism could contribute to the presence of weakly charged black holes contributing to the dark matter in present-day universe \cite{Araya_2023}.

    \item \textbf{Hawking Evaporation and Charge Extremalization}: As a weakly charged black hole undergoes Hawking radiation, it can approach an extremal state where $Q/M \rightarrow 1$. At this stage, theoretical models suggest that an extremal black hole may split into two ``daughter" black holes with $Q_1 < M_1$ and $Q_2 > M_2$, assuming the weak gravity conjecture holds \cite{2007JHEP...06..060A,2007JHEP...12..068K,2018JHEP...10..004C}. This process would also provide a pathway for the formation of charged black-hole binaries.
\end{itemize}

\subsection{Charge Stability and Discharge Mechanisms}

Although any free charge in the universe would generally attract opposite charges and discharge over time, several factors can stabilize a black hole’s charge:

\begin{itemize}
    \item The timescale for charge accretion versus charge discharge is crucial. In many scenarios, the discharge timescale is comparable to the free-fall or dynamical timescale of the system, allowing a black hole to maintain charge for extended periods once they have acquired it \cite{Zajacek_2018}.
    
    \item In cases where rotating black holes are immersed in magnetic fields, difference in electric potential can continuously replenish charge faster than it can neutralize, leading to quasi-stable charge retention.
    
    \item The presence of additional dimensions or non-standard gravitational effects, such as those in higher-dimensional Randall-Sundrum models \cite{1999PhRvL..83.3370R,1999PhRvL..83.4690R} and quadratic gravity \cite{2018PhRvD..98b1502P,2020PhRvD.101b4027P}, induces space-time deformations leading to ``tidal'' charges for non-rotating black holes. Electric charge can effectively mimic these additional black-hole parameters in the extensions of general relativity. 
\end{itemize}

\subsection{Relevance to This Work}

For the purposes of this study, we assume that black holes in binary systems can acquire and retain a weak but stable charge through one or more of the above mechanisms during their lifetime. While the precise astrophysical relevance of each mechanism remains uncertain, the ability of charged black holes to influence particle trapping in binary systems is of key interest. 

The analysis in this paper focuses on cases where the charge-to-mass ratio of each black hole remains small enough that it does not significantly affect the spacetime metric but is sufficient to modify particle dynamics in the vicinity. The charge mechanisms outlined here provide justification for the presence of charge in the context of binary black hole systems. A more detailed discussion of charge stability timescales and their astrophysical constraints can be found in \cite{Zajacek_2018}.

\section{Derivation of the Approximate Hyperbolic Potential Near the Center}\label{ap_sec:potential_derivation}

We consider two identical black holes, each of mass $M$ and charge $Q$, located at $(b,0,0)$ and $(-b,0,0)$ in the $xy$-plane. A test particle of mass $m$ and charge $e$ is placed at $(x_p,y_p,z_p)$ near the origin, with $\sqrt{x_p^2 + y_p^2 + z_p^2} \ll b$. 

\subsection*{1. Newtonian Forces}
Each black hole exerts a force on the particle due to gravitational attraction and electrostatic interaction. Denote:
\begin{equation}
     \mathbf{F}_{1p} = \bigl(F_{1px},\,F_{1py},\,F_{1pz}\bigr), 
    \quad 
    \mathbf{F}_{2p} = \bigl(F_{2px},\,F_{2py},\,F_{2pz}\bigr).
\end{equation}

For black hole 1 (at $x=b,\,y=0,\,z=0$), the $x$-component of the force is:
\begin{equation}
\label{eq:F1px}
F_{1px} 
= 
\frac{GMm - k Q e}{(x_p - b)^2 + y_p^2 + z_p^2}
\,\frac{(x_p - b)}{\sqrt{(x_p - b)^2 + y_p^2 + z_p^2}},
\end{equation}
and similarly for $F_{1py}$ and $F_{1pz}$. An analogous expression holds for BH2 (at $-b,\,0,\,0$):
\begin{equation}
    F_{2px} 
= 
\frac{GMm - k Q e}{(x_p + b)^2 + y_p^2 + z_p^2}
\,\frac{(x_p + b)}{\sqrt{(x_p + b)^2 + y_p^2 + z_p^2}}.
\end{equation}

The net force on the particle is thus:
\begin{equation}
      \mathbf{F}_{\mathrm{net}} 
  = 
  \bigl(F_{1px}+F_{2px},\; F_{1py}+F_{2py},\; F_{1pz}+F_{2pz}\bigr).
\end{equation}

\subsection*{2. Taylor Expansion of the Forces Up to Second Order}

Define the dimensionless small parameters:
\begin{equation}
    \epsilon_x = \frac{x_p}{b}, 
   \quad
   \epsilon_y = \frac{y_p}{b},
   \quad
   \epsilon_z = \frac{z_p}{b},
   \quad\text{with}\quad 
   \sqrt{\epsilon_x^2 + \epsilon_y^2 + \epsilon_z^2} \ll 1.
\end{equation}

We perform a Taylor expansion around $(\epsilon_x, \epsilon_y, \epsilon_z) = (0,0,0)$ (i.e., around $x_p=0,y_p=0,z_p=0$). As an example, consider the denominator in Eq.~\eqref{eq:F1px}:
\begin{equation}
     r_1^2 
  = (x_p - b)^2 + y_p^2 + z_p^2
  = b^2 \bigl( (1 - \epsilon_x)^2 + \epsilon_y^2 + \epsilon_z^2 \bigr).
\end{equation}

Expanding the reciprocal up to second order gives:
\begin{equation}
      \frac{1}{(x_p - b)^2 + y_p^2 + z_p^2}
  \approx
  \frac{1}{b^2}\,\Bigl[\,1 + 2\,\epsilon_x - \epsilon_x^2 - \epsilon_y^2 - \epsilon_z^2 \Bigr].
\end{equation}

A similar expansion is applied to the square root term in the force components.

\paragraph{Example (leading terms):}
For black hole 1 (at $+b$ on the $x$-axis),
\begin{align*}
   & F_{1px}
   \approx
   (GMm - k Q e)\,
   \frac{1}{b^2}
   \Bigl[
       -\,1 
       + 2\,\epsilon_x
       - \epsilon_x^2 
       - \epsilon_y^2 
       - \epsilon_z^2
   \Bigr],
   \\
   & F_{1py} 
   \approx
   (GMm - k Q e)\,
   \frac{1}{b^2}
   \Bigl[
       \; \epsilon_y
   \Bigr],
\end{align*}
where omitted terms represent other second-order corrections and cross-terms. A similar expansion is performed for \(\mathbf{F}_{2p}\) with \(b\) replaced by \(-b\). The net force is obtained by summing contributions from both black holes.

\subsection*{3. Error Estimates for \(\epsilon_x,\epsilon_y,\epsilon_z\)}

Since each expansion is truncated at second order, the leading-order neglected terms are of order:
\begin{equation}
     \mathcal{O}\!\bigl(\epsilon_x^3,\,\epsilon_y^3,\,\epsilon_z^3,\,\epsilon_x\,\epsilon_y^2,\dots\bigr).
\end{equation}

For displacements $|x_p|, |y_p|, |z_p| \leq 0.1b$, we obtain:
\begin{equation}
 |\epsilon_x|, |\epsilon_y|, |\epsilon_z| \leq 0.1,
   \quad
   \epsilon_x^2, \epsilon_y^2, \epsilon_z^2 \leq 0.01.   
\end{equation}
Thus, any neglected third-order terms are at most $< 0.001$, i.e., at most 1\% of the retained second-order terms, justifying truncation at second order.

Furthermore, as discussed in the main text, the smallness of these higher-order terms explains why simulations that included nonlinearities in the gravitational and electric fields still exhibited stable trapping within a radius of $\sim 0.1b$. The results in Appendix B confirm that third-order corrections remain negligible within this region, supporting the numerical evidence for stable confinement.

\subsection*{4. Integrating the Force to Obtain the Potential}

The net force is related to the potential $V(x_p,y_p,z_p)$ by:
\begin{equation}
     F_{\mathrm{net},\,x} = - \frac{\partial V}{\partial x_p}, 
   \quad
   F_{\mathrm{net},\,y} = - \frac{\partial V}{\partial y_p}, 
   \quad
   F_{\mathrm{net},\,z} = - \frac{\partial V}{\partial z_p}.
\end{equation}

Integrating term by term, we obtain:
\begin{equation}
\label{eq:V_approx_saddle}
   V(x_p,y_p,z_p)
   \approx
   \frac{GMm - kQe}{b^3}
   \Bigl(\!
         -\,2\,x_p^2 + y_p^2 + z_p^2
   \Bigr).
\end{equation}

The key point is that, due to the two identical sources at $\pm b$, the coefficient in front of $x_p^2$ differs from that of $y_p^2$ and $z_p^2$, giving the characteristic "saddle shape" in the $x$-direction. The potential \eqref{eq:V_approx_saddle} is a valid approximation within the stable trapping region.

\section{Coupled Mathieu‐Type Equations and Their Analytical Solution}\label{ap_sec:stab_eq_derivation}

We begin with the pair of coupled equations governing the motion of the test particle, parametrized by dimensionless time \(\tau\):

\begin{equation}\label{eq:mat_x}
  \frac{d^2 x}{d\tau^2} + a\,x 
  + 2\,q\,\Bigl[\cos(2\tau)\,x + \sin(2\tau)\,y\Bigr] 
  = 0,
\end{equation}

\begin{equation}\label{eq:mat_y}
  \frac{d^2 y}{d\tau^2} + a\,y 
  - 2\,q\,\Bigl[\cos(2\tau)\,y - \sin(2\tau)\,x\Bigr] 
  = 0.
\end{equation}

To simplify the analysis, we introduce the complex variable:

\begin{equation}
  z(\tau) = x(\tau) + i\,y(\tau).    
\end{equation}

Adding Eq.~\eqref{eq:mat_x} and \(i\) times Eq.~\eqref{eq:mat_y} yields a single equation for \(z(\tau)\):

\begin{equation}\label{eq:z_eq}
  \frac{d^2 z}{d\tau^2} + a\,z 
  + 2\,q\,z^*(\tau)\,\mathrm{e}^{\,i\,2\tau}
  = 0,
\end{equation}

where \(z^*(\tau)\) is the complex conjugate of \(z(\tau)\).  

\subsection*{Rotating‐Wave Approximation}

We now impose the \emph{rotating‐wave approximation} by writing

\begin{equation}\label{eq:z_ansatz}
  z(\tau) = f(\tau)\,\mathrm{e}^{\,i\tau},
  \quad
  \overline{z}(\tau) = f^*(\tau)\,\mathrm{e}^{-\,i\tau}.
\end{equation}

Differentiating yields:

\begin{equation}
  \frac{d z}{d\tau}
  =
  \mathrm{e}^{\,i\tau}\,\bigl(f'(\tau) + i\,f(\tau)\bigr),
  \qquad
  \frac{d^2 z}{d\tau^2}
  =
  \mathrm{e}^{\,i\tau}
  \Bigl[
    f''(\tau) + 2\,i\,f'(\tau) - f(\tau)
  \Bigr],
\end{equation}

where primes denote derivatives with respect to \(\tau\). Substituting these into Eq.~\eqref{eq:z_eq} and neglecting high-frequency counterrotating terms simplifies the equation. The resulting expression allows us to eliminate any dependence on \(f^*(\tau)\), leading to a \textbf{fourth‐order} ordinary differential equation for \(f(\tau)\):

\begin{equation}\label{eq:f_fourth_order}
  \frac{d^4 f}{d\tau^4}
  + 2\,(1 + a)\,\frac{d^2 f}{d\tau^2}
  + \Bigl[(1 - a)^2 - 4\,q^2\Bigr]\,f(\tau)
  = 0.
\end{equation}

\subsection*{Solving the Fourth‐Order ODE and Stability}

Since Eq.~\eqref{eq:f_fourth_order} has constant coefficients, we assume an exponential solution:

\begin{equation}
  f(\tau) = A\,\mathrm{e}^{\,\beta\,\tau}.
\end{equation}

Substituting this ansatz into Eq.~\eqref{eq:f_fourth_order} gives the characteristic equation:

\begin{equation}\label{eq:beta_char}
  \beta^4
  + 2\,(1 + a)\,\beta^2
  + (1 - a)^2 - 4\,q^2
  = 0.
\end{equation}

Defining \(\beta^2 = X\) reduces Eq.~\eqref{eq:beta_char} to a quadratic in \(X\). The condition for "stability" (i.e., bounded, oscillatory solutions) is that all roots \(\beta\) remain purely imaginary, ensuring \(\mathrm{Re}(\beta) = 0\). This leads to the key "stability criterion":

\begin{equation}
  a \geq 1 + 2\,q
  \quad\text{or}\quad
  a \in \bigl[-q^2,\,1 - 2\,q\bigr].
\end{equation}

These conditions ensure that all solutions remain oscillatory rather than exhibiting exponential runaway behavior, thereby confirming the stability of particle motion in this system.

\section{Stability Parameters in Terms of Fundamental Physical Quantities}

The stability parameters $a$ and $q$ remain unchanged under a simultaneous scaling of black hole mass $M$ and charge $Q$. The Lorentz term in the equations of motion (see Section 3 of the main text) is also unaffected when the magnetic field $B$ is scaled appropriately. This scaling ensures that the system exhibits identical dynamical behavior, as discussed in Section 2.1.

\subsection{Stability Parameters and Their Dependence on Fundamental Constants}

The stability parameters $a$ and $q$ are originally defined as:

\begin{equation}
    a = -\frac{K(\alpha - 1)}{m\Omega^2}, \quad q = \frac{K(\alpha + 1)}{2m\Omega^2}.
\end{equation}

where $K$ is a proportionality constant associated with the gravitational and electrostatic forces. Using Kepler’s third law:

\begin{equation}
    \frac{K}{m\Omega^2} = 4\left(1 - \frac{kQe}{GMm}\right),
\end{equation}

we express $a$ and $q$ in terms of fundamental constants:

\begin{equation}
    a = -4\left(1 - \frac{kQe}{GMm} \right) \cdot (\alpha - 1), \quad q = \frac{4}{2} \left(1 - \frac{kQe}{GMm} \right) \cdot (\alpha + 1).
\end{equation}

Here, $k$ is the Coulomb constant, $e$ is the charge of the test particle, $Q$ is the charge of the black holes, $G$ is the gravitational constant, $M$ is the black hole mass, and $m$ is the test particle mass.

\subsection{Invariance of Stability Parameters Under \( M, Q \) Scaling}

If both $M$ and $Q$ are scaled by the same factor, say $M' = \lambda M$ and $Q' = \lambda Q$, then the ratio $Q/M$ remains unchanged, i.e., $Q'/M' = Q/M$. 

Since $a$ and $q$ depend only on the factor $Q/M$ and not on the absolute values of $M$ or $Q$, it follows that:

\begin{equation}
    a' = a, \quad q' = q.
\end{equation}

Thus, under this rescaling, the system remains within the same stability region in the $(q, a)$ diagram.

\subsection{Scaling the Magnetic Field to Maintain the Same Lorentz Force}

The equations of motion for the system include a Lorentz force term given by:

\begin{equation}
    \frac{d^2 x}{d\tau^2} - \left(a + 2q \cos(2\tau)\right) x - 2q \sin(2\tau) y + \frac{\omega_o}{\Omega} \frac{dy}{d\tau} = 0,
\end{equation}

\begin{equation}
    \frac{d^2 y}{d\tau^2} - \left(a - 2q \cos(2\tau)\right) y - 2q \sin(2\tau) x - \frac{\omega_o}{\Omega} \frac{dx}{d\tau} = 0,
\end{equation}

where $\omega_o = eB/m$ is the cyclotron frequency.

Since $\Omega$ (the black hole orbital frequency) scales as:

\begin{equation}
    \Omega' = \Omega \lambda^{-1/2},
\end{equation}

we require $\omega_o / \Omega$ to remain unchanged. This means the magnetic field should scale as:

\begin{equation}
    B' = B \cdot \lambda^{1/2}.
\end{equation}

By increasing both the black hole mass and charge by a factor of $\lambda = 10$, and simultaneously scaling the magnetic field by $\sqrt{10}$, we preserve the form of the equations of motion, ensuring that the system’s dynamics remain unchanged. Consequently, the test particle follows an identical trajectory, confirming the theoretical consistency of our results. This explains why numerical simulations yield the same particle trajectories when $M$ and $Q$ are scaled together, provided the magnetic field is adjusted accordingly.



\begin{figure}[h!]
    \label{fig:sim2}
\end{figure}

\end{document}